\documentclass[aps,prb, amsfonts, amssymb, amsmath, superscriptaddress, notitlepage,twocolumn,10pt,nobalancelastpage]{revtex4-2}
\usepackage{graphicx} 
\usepackage{hyperref}
\usepackage{bm}
\hypersetup{ 
    colorlinks=true, %set true if you want colored links
    linktoc=all,     %set to all if you want both sections and subsections linked
    linkcolor=blue,  %choose some color if you want links to stand out
    citecolor =blue
} 
\usepackage{amsmath}

\newcommand{\bk}{\mathbf{k}}

\newcommand{\br}{\mathbf{r}}

\newcommand{\bK}{\mathbf{K}}

\newcommand{\ba}{\mathbf{a}}

\newcommand{\nn}{\nonumber}

\begin{document}

\title{Majorana Fermion Mean-Field Theories of Kitaev Quantum Spin Liquids}

\author{Shahnam Ghanbari Saheli}
\affiliation{London Centre for Nanotechnology, University College London, Gordon St., London, WC1H 0AH, United Kingdom}
\author{Jennifer Lin}
\affiliation{London Centre for Nanotechnology, University College London, Gordon St., London, WC1H 0AH, United Kingdom}
\author{Huanzhi Hu}
\affiliation{London Centre for Nanotechnology, University College London, Gordon St., London, WC1H 0AH, United Kingdom}
\author{Frank Kr\"uger}
\affiliation{London Centre for Nanotechnology, University College London, Gordon St., London, WC1H 0AH, United Kingdom}
\affiliation{ISIS Facility, Rutherford Appleton Laboratory, Chilton, Didcot, Oxfordshire OX11 0QX, United Kingdom}

\begin{abstract}
We determine the phase diagrams of anisotropic Kitaev-Heisenberg models on the honeycomb lattice using parton mean-field theories based on different Majorana 
fermion representations of the $S=1/2$ spin operators. Firstly, we use a two-dimensional Jordan-Wigner transformation (JWT) involving a semi-infinite snake string operator. 
In order to ensure that the fermionized Hamiltonian remains local we consider the limit of extreme Ising exchange anisotropy in the Heisenberg sector. Secondly, we use the 
conventional Kitaev representation in terms of four Majorana fermions subject to local constraints, which we enforce through Lagrange multipliers. For both representations we 
self-consistently decouple the interaction terms in the bond and magnetization channels and determine the phase diagrams as a function of the anisotropy of the Kitaev couplings 
and the relative strength of the Ising exchange. While both mean-field theories produce identical phase boundaries for the topological phase transition between the gapless and 
gapped Kitaev quantum spin liquids, the JWT fails to correctly describe the the magnetic instability and finite-temperature behavior. Our results show that the magnetic phase 
transition is first order at low temperatures but becomes continuous above a certain temperature. At this energy scale we also observe a finite temperature crossover 
on the quantum-spin-liquid side, from a fractionalized paramagnet at low temperatures, in which gapped flux excitations are frozen out,  to a conventional paramagnet 
at high temperatures.

\end{abstract}

\maketitle

\section{Introduction}

Quantum Spin Liquids (QSLs) \cite{savary+16,knolle+18,hermanns+18} are a novel class of materials in which geometric and/or exchange frustration suppresses 
magnetic order down to absolute zero temperature. Because of the topological character of the ground-state wave function 
with a special type of long-range quantum entanglement, QSLs exhibit exotic fractional excitations \cite{Broholm+20}, which are believed 
to hold great potential for quantum communication and computation \cite{Wen+19}. These concepts were put on a firm footing in the 
seminal work by Alexei Kitaev \cite{Kitaev06} who constructed an exactly solvable QSL model on the honeycomb lattice and demonstrated 
that the spins break-up (fractionalize) into a set of Majorana fermions. The emergent fermions essentially behave as the electrons 
in graphene with a relativistic Dirac dispersion, although they don’t carry electric charge and are coupled to gauge fields. 

Although the bond-directional dependence of the Ising exchange anisotropy in the Kitaev model might seem  artificial,  it was later realized  that, as a result of spin-orbital 
entanglement \cite{Jackeli+09},   the Kitaev couplings can play a dominant role in honeycomb Iridates  and Ruthenates, such as  
Na$_2$IrO$_3$ \cite{Singh+10,Liu+11,Singh+12,Choi+12,Ye+12}, 
$\alpha$-Li$_2$IrO$_3$ \cite{Singh+12}, $\beta$-Li$_2$IrO$_3$ \cite{Takayama+15}, $\gamma$-Li$_2$IrO$_3$ \cite{Modic+14} and 
$\alpha$-RuCl$_3$ \cite{Plumb+14,Banerjee+15,Banerjee+17}.  However, small additional magnetic interactions such as Heisenberg terms 
drive these systems into a magnetically ordered state that forms at  low temperatures. Nevertheless, at higher temperatures or in applied magnetic 
field signatures of the nearby Kitaev QSL state are seen \cite{Takagi+19}.

While the theoretical interest in the novel fundamental physics of QSLs is  considerable,  the experimental identification of 
QSLs has proven difficult. While the emergent fermions have manifestations in specific heat and thermal transport properties \cite{Takagi+19,Go+19,Yokoi+21}, 
rather indirect evidence comes from the lack of magnetic ordering seen in NMR, $\mu$SR and neutron diffraction, as well as from the 
absence of sharp quasiparticle excitations in neutron scattering. Unlike in the case of Heisenberg spin-1/2 chain 
systems where the measured intensity variation is quantitatively understood from the continuum of fractionalized spinon 
excitations \cite{Mourigal+13,Lake+13}, in the case of two-dimensional QSLs theoretical techniques are yet to be developed to quantitatively 
understand finite temperature excitation spectra.

A possible way to distinguish signatures of fractionalization from diffuse scattering originating from disorder or short ranged and lived 
quasiparticle excitations is through entanglement witnesses such as quantum Fisher information \cite{Hauke+16} which can be directly computed from the 
 dynamic susceptibilities measured in inelastic neutron scattering experiments \cite{Scheie+21}. In the case of the idealized Kitaev model it was demonstrated 
 theoretically \cite{Knolle+14,Knolle+15}  that the magnetic structure factor shows  signatures of fractionalized Majorana fermions and fluxes of $Z_2$ gauge fields
 that are in qualitative agreement with the finite-temperature excitation spectrum of $\alpha$-RuCl$_3$ \cite{Banerjee+15,Banerjee+17}.
 More recently, the theoretical approach was extended beyond the integrable point of the pure Kitaev model, using an augmented 
parton mean-field theory based on the Kitaev Majorana representation \cite{Knolle+18b}. Finally, by combining the density-matrix renormalization (DMRG) ground 
state method and a matrix-product state (MPS) based dynamical algorithm \cite{Gohlke+17} it was demonstrated that the spectra of the Kitaev-Heisenberg model 
close to the QSL phase show proximate spin-liquid features.

Although the physics of the Kitaev model is naturally captured in terms of Majorana fermions, phase diagrams of the Kitaev-Heisenberg model and extensions thereof
were calculated in terms of complex spin-1/2 fermionic spinons, either on the level of SU(2) slave fermion mean-field theory \cite{Burnell+11,schaffer+12} or 
numerically by means of pseudo-fermion functional renormalization group \cite{Singh+12,Jiang+11,Reuther+11}.

In this article we compute zero and finite temperature mean-field phase diagrams in terms of fractionalized  Majorana fermion degrees of freedom. There exist different   
type of representations of the spin-1/2 operators in terms of Majorana fermions which are equivalent for the description of the ground-state properties of the 
pure isotropic or anisotropic Kitaev model \cite{Chen+08,Fu+18}, but not necessarily if finite-temperature excitations are considered or additional interactions are 
taken into account. We focus on two representations, the one originally introduced by Kitaev \cite{Kitaev06} and the two-dimensional Jordan-Wigner transformation (JWT)
\cite{Feng+07,Chen+07,Chen+08,Dora+18}.

In addition to Kitaev and Heisenberg exchange we will consider magnetic exchange anisotropy, which as a result of the directional dependence of the Kitaev coupling 
induces spatial anisotropy. For the pure Kitaev model anisotropy is known to result in a topological phase transition \cite{Kitaev06}  from a QSL hosting gapless Majorana and 
gapped flux excitations to a gapped $Z_2$ one with Abelian excitations \cite{Gohlke+17}. 

The outline of the paper is as follows. In Sec.~\ref{sec.model} we define and motivate the Hamiltonian of the anisotropic $S=1/2$ Kitaev-Heisenberg model on the 
honeycomb lattice. The two-dimensional JWT and consecutive mean-field decoupling scheme are introduced in Sec.~\ref{sec.JWT}, where the underlying string operator 
is defined such that the fermionized Hamiltonian remains local in the extreme Ising limit of the Heisenberg exchange interaction.  In Sec.~\ref{sec.Kitaev} we map the 
spin Hamiltonian to a set of four Majorana fermions, following the original construction by Kitaev, and enforce the Hilbert space constraint through a Lagrange multiplier. 
We discuss the mean-field decoupling of the interaction terms in bond and magnetization channels and determine the Lagrange multiplier as a function of 
the mean-field parameters. 

Our results are presented in Sec.~\ref{sec.results}. We first demonstrate that the two mean-field theories result in identical phase boundaries for
the topological transition between the gapless and gapped Kitaev QSLs. Interestingly, the anisotropy of the Kitaev coupling and the Ising exchange cooperate in driving the 
transition.We then determine the antiferromagnetic instability driven by the Ising exchange. This transition is strongly first order and not correctly described by the JWT 
mean-field theory.  We finally 
 determine the finite-temperature phase diagram and show that the magnetic phase transition becomes continuous above a certain temperature.  At this temperature scale, 
 the specific heat above the Kitaev QSL shows a peak, indicating a crossover between a fractionalized paramagnet with frozen $Z_2$ flux excitations to a conventional 
 paramagnetic state at higher temperatures.  In Sec.~\ref{sec.discussion}
  we summarize and discuss our results.

%%%%%%%%%%%%%%

\section{Model}
\label{sec.model}

Our starting model is the anisotropic $S=1/2$ Kitaev-Heisenberg model on the honeycomb lattice in the limit of an extreme Ising anisotropy in the Heisenberg
sector. The Hamiltonian of the model is given by
\begin{equation}
\label{eq.Ham_spin}
\hat {\cal H} = \sum_{\gamma = x,y,z}  \sum_{\langle i,j \rangle_\gamma} K_\gamma \hat \sigma_i^\gamma \hat \sigma_j^\gamma + J \sum_{\langle i,j\rangle} \hat \sigma_i^z \hat \sigma_j^z,
\end{equation}
where $\hat\sigma^\gamma$ are the spin-1/2 operators in units of $\hbar/2$, $\hat S^\gamma = \frac{\hbar}{2} \hat\sigma^\gamma$, satisfying 
the spin commutator relations $[\hat\sigma_i^\alpha,\hat\sigma_j^\beta] = 2 \delta_{ij}\epsilon_{\alpha\beta\gamma}\hat\sigma_i^\gamma$.

\begin{figure}[t]
 \includegraphics[width=\linewidth]{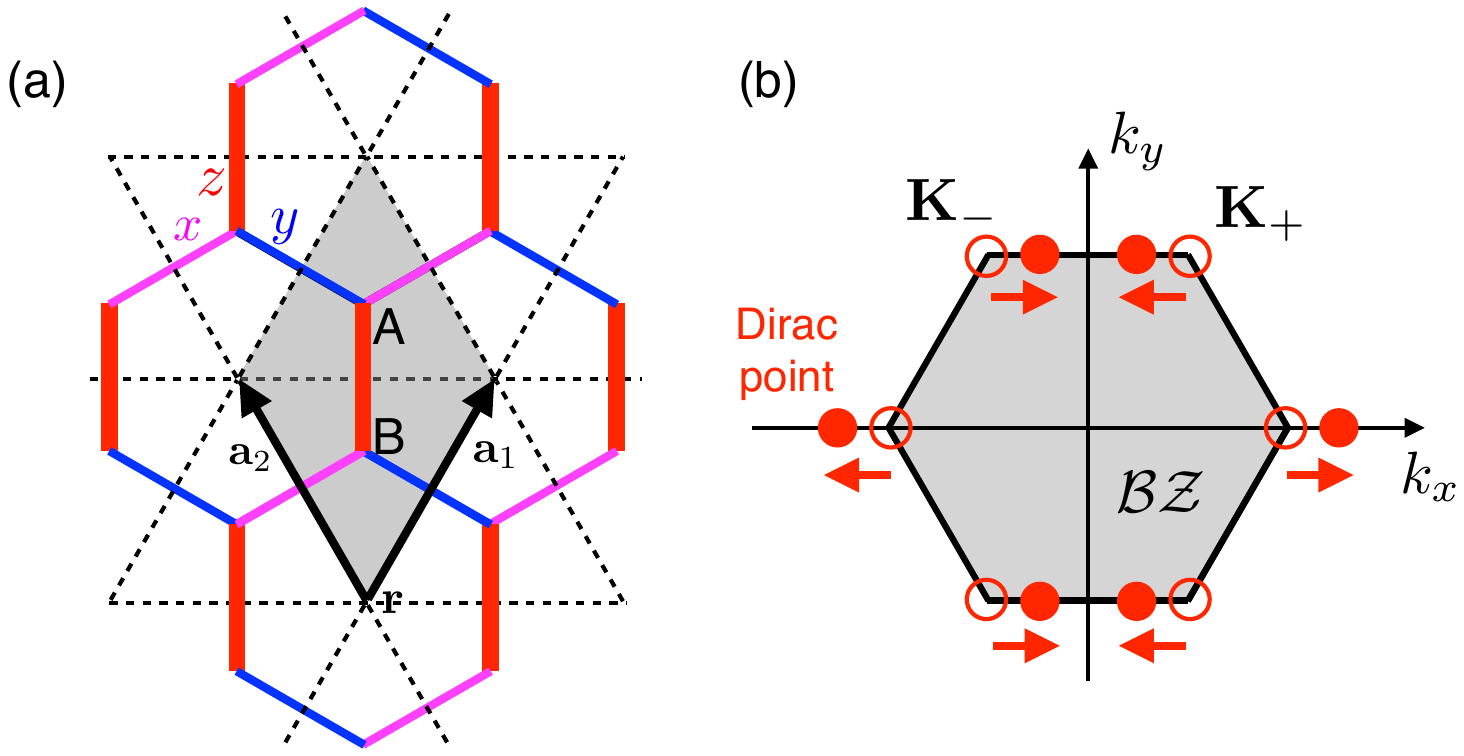}
 \caption{(a) Illustration of the Kitaev model on the honeycomb lattice. The three inequivalent nearest neighbor bonds are labelled by $\gamma=x,y,z$ and shown in different colors. Along a bond in 
  the $\gamma$ direction only the spin components $S^\gamma$ between the neighboring sites are coupled. We allow the Kitaev couplings to be anisotropic and include an additions Ising exchange 
  between spin-$z$ components on all nearest neighbor bonds. The unit cell of the honeycomb lattice, shaded in grey,  contains two lattice sites labelled by $A$ and $B$.  (b) The pure Kitaev model is 
  exactly solvable in terms of Majorana fermions. In the isotropic limit one band is gapless with Dirac points at the corners $\bK_+$ and $\bK_-$ of the hexagonal Brillouin zone ($\mathcal{BZ}$). 
  With increasing anisotropy $\delta$ and Ising exchange $\alpha$ the Dirac points move along the edges of the $\mathcal{BZ}$ and eventually merge, corresponding to a topological phase transition 
  to a gapped Kitaev QSL. Sufficiently strong $\alpha$ results in a first-order transition to an antiferromagnet with fully gapped Majorana fermion spectrum.}
\label{figure1}
\end{figure}

The Kitaev couplings $K_\gamma$ are illustrated in Fig.~\ref{figure1}(a). Along each of the three inequivalent nearest neighbor bonds, labelled by $\gamma=x,y,z$, different 
spin components are coupled, e.g. along the $x$-bonds the Kitaev coupling is $K_x  \hat \sigma_i^x \hat \sigma_j^x$. In this work we consider antiferromagnetic Kitaev couplings and allow the 
coupling between spin-$z$ components to be stronger than those between the $x$ and $y$ components, $K_z \ge K_x =K_y =K >0$. Because of the bond-directional nature of the Kitaev coupling, 
this spin-exchange anisotropy is linked to a strong spatial anisotropy.  

By symmetry one should also expect spin-exchange anisotropy in the Heisenberg interactions. For reasons that we will explain later we focus on the case of very strong Ising anisotropy, 
$J=J_z>0$ and $J_x=J_y=0$.

The zero-temperature phase diagram of the model is controlled by the two dimensionless parameters
\begin{equation}
\label{eq.couplings}
\alpha = \frac{J}{K}\quad\textrm{and}\quad \delta=\frac{K_z - K}{K}.
\end{equation}

For $\alpha=0$ the model reduces to an anisotropic Kitaev model which is exactly solvable in terms of Majorana fermions, either
by using the original Kitaev construction \cite{Kitaev06} or a two-dimensional Jordan-Wigner transformation \cite{Chen+08}. In both cases one obtains flat bands, corresponding to local flux excitations, 
and a gapless dispersive band with Dirac points at the Fermi level. In the isotropic Kitaev model ($\delta=0$) the Dirac points are located at the corners $\bK_\pm=2\pi(\pm1/(3\sqrt{3},1/3)$ 
of the hexagonal Brillouin zone ($\mathcal{BZ}$) [see Fig.~\ref{figure1}(b)]. Anisotropy in the Kitaev couplings is known to drive a topological 
phase transition from a gapless to a gapped QSL \cite{Kitaev06}. With Increasing anisotropy $\delta$, the Dirac points move along the edges of the 
$\mathcal{BZ}$ and merge when $\delta_c=1$, corresponding to $K_z/K=2$. At this point the dispersive Majorana band exhibits a semi-Dirac point at $2\pi(0,1/3)$, 
which is a quadratic band touching point along the edge but relativistic in the transverse direction. For values $\delta>1$ the excitations become gapped. This behavior is very similar to the 
topological phase transition proposed to occur for electrons moving in strained honeycomb lattices \cite{Dietl+08,Montambaux+09,Banerjee+09,Uryszek+19,Uryszek+20} and observed experimentally 
in black phosphorus \cite{Kim+15,Kim+17}.

We will see that $\alpha = J/K$ has a similar effect on the Dirac band and cooperates with the anisotropy in driving the topological phase transition. In addition, the  flux excitations 
become weakly dispersive for $\alpha>0$. As one might expect, sufficiently strong $\alpha$ leads to a first-order transition between a Kitaev QSL and an Ising antiferromagnet 
with a gapped Majorana fermion spectrum of strongly hybridized bands.

\section{Jordan-Wigner Transformation}
\label{sec.JWT}

The Jordan-Wigner transformation (JWT) is usually used to express one-dimensional $S=1/2$ spin Hamiltonians in terms of 
spinless fermions with creation and annihilation operators $\hat d^\dagger_n$, $\hat d_n$, where $n$ labels the site along the 
one-dimensional lattice. It is natural to identify the no-fermions state $|0\rangle$ with the eigenstate $|\uparrow\rangle$ of the $\hat \sigma^z$
spin operator and the singly occupied state $|1\rangle$ with $|\downarrow\rangle$. However, since spin operators on different sites commute 
while fermionic operators anti-commute, it is not possible to define a local transformation. Instead one needs to include a semi-infinite 
string operator 
\begin{equation}
\hat D_n = \prod_{\ell<n} (1-2\hat d^\dagger_\ell \hat d_\ell).
\end{equation}
to match the quantum statistics of spins and fermions and define the one-dimensional JWT as 
\begin{eqnarray}
 \hat \sigma^z_n &  = &  1-2 \hat d^\dagger_n \hat d_n =  (\hat d^\dagger_n+\hat d_n)(\hat d^\dagger_n-\hat d_n),\\
 \hat \sigma^x_n & = &    \hat D_n (\hat d^\dagger_n+\hat d_n),\\
 \hat \sigma^y_n & = &    i\hat D_n (\hat d^\dagger_n-\hat d_n).
\end{eqnarray}
It is easy to check that the string operator is hermitian, $\hat D^\dagger_n = \hat D_n$, and satisfies $\hat D^2_n =1$,  
$\hat D_n \hat D_{n+1} = 1 - 2 \hat d^\dagger_n \hat d_n$ and 
 $[\hat d^\dagger_n,\hat D_n] = [\hat d_n,\hat D_n] = 0$.
 
 \begin{figure}[t]
 \includegraphics[width=0.6\linewidth]{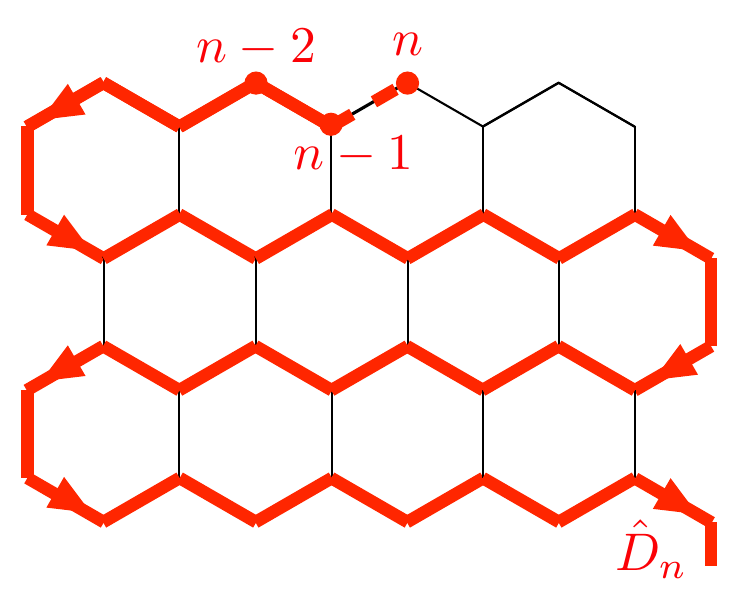}
 \caption{Illustration of the snake string operators used in the two-dimensional Jordan-Wigner transformation (JWT) of the Kitaev-Ising model.}
\label{figure2}
\end{figure}

 These properties of the string operator ensure that one-dimensional spin Hamiltonians with short-ranged spin interactions 
 remain short ranged after JWT. The generalization of the JWT to two dimensions is problematic for several reasons. 
 Firstly, the string operator connecting a given lattice site to infinity is not uniquely defined, and in principle gauge transformations 
 corresponding to  deformations of the string need to be taken into account \cite{Li+22}. Secondly, nearest neighbor sites in the two dimensional 
 lattice are not necessarily nearest neighbors along the string. As a result, the fermionized Hamiltonian will contain non-local interactions
 involving segments of string operators. 
 
 The Kitaev model on the honeycomb lattice is an example where the second problem of non-locality can be circumvented by defining 
 snake string operators \cite{Feng+07,Chen+07,Chen+08} shown in Fig.~\ref{figure2}. In this case, the $x$ and $y$ bonds, which involve the string 
 operators, couple nearest neighbors along the string. Using that $\hat D_n \hat D_{n+1} = 1 - 2 \hat d^\dagger_n \hat d_n$ we obtain
 \begin{eqnarray}
 \hat \sigma^x_n \hat \sigma^x_{n+1} & = & (\hat d^\dagger_n-\hat d_n) (\hat d^\dagger_{n+1}+\hat d_{n+1}),\\
 \hat \sigma^y_n \hat \sigma^y_{n+1} & = & (\hat d^\dagger_{n+1}-\hat d_{n+1}) (\hat d^\dagger_n+\hat d_n).
 \end{eqnarray}
Note that the $z$ bonds connect spins that are not nearest neighbors along the snake string. As a result, any Hamiltonian that 
involves couplings between the $x$ or $y$ spin components along the $z$ bonds, e.g. the Kitaev-Heisenberg model, would be non-local in 
terms of the Jordan-Wigner fermions. This however is not the case for the pure Kitaev model or for our model with additional Ising couplings 
$\hat \sigma^z_i \hat \sigma^z_j$ on all nearest-neighbor bonds.

The resulting Hamiltonian of the Kitaev-Ising model is given by
\begin{eqnarray}
\hat {\cal H}/K  &  = & \sum_\br \sum_{i=1,2} (\hat d^\dagger_{A,\br}-\hat d_{A,\br}) (\hat d^\dagger_{B,\br+\ba_i}+\hat d_{B,\br+\ba_i})\nn\\
& &  +\alpha  \sum_\br \sum_{i=1,2} (\hat d^\dagger_{A,\br}+\hat d_{A,\br}) (\hat d^\dagger_{A,\br}-\hat d_{A,\br})\nn\\
& & \quad\times (\hat d^\dagger_{B,\br+\ba_i}+\hat d_{B,\br+\ba_i}) (\hat d^\dagger_{B,\br+\ba_i}-\hat d_{B,\br+\ba_i})\nn\\
& & +(1+\delta+\alpha) \sum_\br  (\hat d^\dagger_{A,\br}+\hat d_{A,\br}) (\hat d^\dagger_{A,\br}-\hat d_{A,\br})\nn\\
& & \quad\times (\hat d^\dagger_{B,\br}+\hat d_{B,\br}) (\hat d^\dagger_{B,\br}-\hat d_{B,\br}),
\end{eqnarray}
where $(\alpha,\br)$ denote the sites of the two-dimensional honeycomb lattice, with $\br$ the unit cell spanned by $\ba_1$ and $\ba_2$ and 
 $\alpha=A,B$ the atom in the unit cell, as illustrated in Fig.~\ref{figure1}(a). The dimensionless coupling constants $\alpha$ and $\delta$ are 
 defined in Eq.~(\ref{eq.couplings}). The interaction terms, which arise from the $\hat \sigma^z_i \hat \sigma^z_j$ terms in the spin Hamiltonian, 
are of strength $J$ along the $x$ and $y$ bonds and of strength $K_z+J$ along the $z$ bonds. 

The Hamiltonian is naturally expressed in terms of Majorana fermions,
\begin{eqnarray}
\hat c_A(\br) & = &  i[\hat d^\dagger_A(\br)-\hat d_A(\br)],\nn\\
\hat \eta^z_A(\br) & = &  \hat d^\dagger_A(\br)+\hat d_A(\br),
\end{eqnarray}
on sub-lattice $A$ and
\begin{eqnarray}
\hat c_B(\br) & = &  \hat d^\dagger_B(\br)+\hat d_B(\br),\nn\\
\hat \eta^z_B(\br) & = & i[ \hat d^\dagger_B(\br)-\hat d_B(\br)],
\end{eqnarray}
on sub-lattice $B$. The meaning of the superscript $z$ on the Majorana fermion $\eta$ will become clear when we compare with the mean-field theory based on the 
Kitaev representation of the spin operators in terms of Majorana fermions. The resulting Hamiltonian is given by 
\begin{eqnarray}
\hat {\cal H}/K  &  = & -i \sum_\br \sum_{i=1,2} \hat c_A (\br) \hat c_B(\br+\ba_i)\nn\\
& &  +\alpha  \sum_\br \sum_{i=1,2} \hat c_A(\br)\hat\eta^z_A(\br) \hat c_B(\br+\ba_i)\hat\eta^z_B(\br+\ba_i) \nn\\
& & +(1+\delta+\alpha) \sum_\br  \hat c_A(\br)\hat\eta^z_A(\br) \hat c_B(\br)\hat\eta^z_B(\br).
\end{eqnarray}

The Majorana operators satisfy $\hat c^\dagger_\sigma(\br) = \hat c_\sigma(\br)$. $(\hat \eta^z_\sigma(\br))^\dagger= \hat \eta^z_\sigma(\br)$ and the anti-commutator relations 
$\{ \hat c_\sigma(\br),\hat c_{\sigma'}(\br')\} = \{ \hat \eta^z_\sigma(\br),\hat \eta^z_{\sigma'}(\br')\} = 2\delta_{\sigma,\sigma'}\delta_{\br,\br'}$ and $\{ \hat c_\sigma(\br),\hat \eta^z_{\sigma'}(\br')\} = 0$.

\subsection{Mean-Field Theory}

We perform a self-consistent mean-field decoupling of the interactions in both the bond and density channels. The former is required 
to recover the physics of the Kitaev model. We define the averages 
\begin{eqnarray}
a_{\br,\br'} & = & -i \left\langle  \hat \eta^z_A (\br) \hat\eta^z_B(\br')  \right\rangle \\
b_{\br,\br'} & = &   -i \left\langle  \hat c_A (\br) \hat c_B(\br')  \right\rangle,
\end{eqnarray} 

where $a_\perp = a_{\br,\br+\ba_i}$ and  $b_\perp = b_{\br,\br+\ba_i}$ for the $x$ and $y$ bonds and 
$a_z = a_{\br,\br}$ and $b_z = b_{\br,\br}$ for the $z$ bonds.  The local staggered magnetization of the antiferromagnetic state is given by  
\begin{equation}
m = i \left\langle   \hat c_\sigma(\br)\hat\eta^z_\sigma (\br) \right\rangle.
\end{equation}
Note that this is indeed the staggered magnetization since the roles of $\hat \eta^z$ and $\hat c$ are switched between the two sub-lattices. 
All mean-field parameters, $a_\perp$, $a_z$, $b_\perp$, $b_z$ and $m$, are real since the corresponding operators are hermitian. 
After Fourier transform, 
\begin{eqnarray}
\label{eq.FT1}
\hat c_\sigma(\br) & = &  \frac{1}{\sqrt{2N}}\sum_\bk \left\{e^{i\bk\br} \hat c_\sigma^\dagger (\bk) + e^{-i\bk\br}\hat c_\sigma(\bk)   \right\},\\
\label{eq.FT2}
\hat \eta_\sigma^z(\br) & = &  \frac{1}{\sqrt{2N}}\sum_\bk \left\{e^{i\bk\br} (\hat \eta^z_\sigma)^\dagger (\bk) + e^{-i\bk\br}\hat \eta_\sigma^z(\bk)   \right\},
\end{eqnarray}
where $N$ denotes the number of unit cells and the momenta $\bk$ are from the hexagonal Brillouin zone ($\cal{BZ}$) shown in Fig.~\ref{figure2}(a), 
the resulting mean-field Hamiltonian in momentum space is given by
\begin{eqnarray}
\frac{\hat{\cal H}_\textrm{mf}}{N K} & = &  \frac{i}{N} \sum_\bk \hat\Psi^\dagger_\bk
 \left(\begin{array}{cccc}0 & -\gamma_c^*& - M & 0 \\ \gamma_c& 0 & 0 & - M \\  M & 0 & 0 & -\gamma^*_z \\ 0 &  M & \gamma_z & 0  \end{array} \right)
\hat \Psi_\bk\\
& & -(1+\delta+\alpha) a_z b_z  -2 \alpha a_\perp b_\perp + (1+\delta+3\alpha) m^2.\nn
\end{eqnarray}
Here $\hat\Psi_\bk = (\hat c_A(\bk),\hat c_B(\bk),\hat \eta_A^z(\bk), \hat \eta_B^z(\bk))^T$ and $M=(1+\delta+3\alpha)m$, 
for brevity, and we have defined the complex valued functions
\begin{eqnarray}
\label{eq_gammac}
\gamma_c (\bk) & = &  (1+\delta+\alpha) a_z + (1+\alpha a_\perp) \left( e^{i\bk\ba_1}+e^{i\bk\ba_2}  \right),\\
\label{eq_gammaeta}
\gamma_z (\bk) & = &  (1+\delta+\alpha) b_z + \alpha b_\perp \left( e^{i\bk\ba_1}+e^{i\bk\ba_2}  \right).
\end{eqnarray}  

The energy eigenvalues of the mean-field Hamiltonian are given by (in units of the Kitaev coupling $K$)
\begin{eqnarray}
\label{eq_dispJWT}
\epsilon_{1,2}^2(\bk) & = &  \frac{|\gamma_c|^2+|\gamma_z|^2}{2}+M^2\nn\\
& & \pm \sqrt{\left( \frac{|\gamma_c|^2-|\gamma_z|^2}{2}  \right)^2+|\gamma_c+\gamma_z|^2  M^2},
\end{eqnarray}
resulting in the free energy density 
\begin{eqnarray}
f & = & -t \sum_{n=1,2}\sum_{\sigma=\pm1}\int_\bk \ln\left(e^{-\sigma|\epsilon_n(\bk)|/t}+1  \right)\\
& & -(1+\delta+\alpha) a_z b_z  -2 \alpha a_\perp b_\perp + (1+\delta+3\alpha)m^2,\nn
\end{eqnarray}
with $t=T/K$ the dimensionless temperature and 
\begin{equation}
\int_\bk\ldots = \frac{1}{V_{\cal{BZ}}} \int_{\cal{BZ}} d^2\bk\ldots,
\end{equation}
for brevity, where $V_{\cal{BZ}}$ denotes the volume of the hexagonal Brillouin zone.

Minimizing the free-energy density $f$ with respect to the mean-field parameters $\xi\in\{a_z, b_z,  a_\perp, b_\perp, m  \}$, $\partial_\xi f=0$,  
we obtain the self-consistency equations
\begin{eqnarray}
a_z  & = &  - \frac{1}{1+\delta+\alpha} \int_\bk F_{b_z}(\bm{\xi},\bk),\\
b_z  & = &  - \frac{1}{1+\delta+\alpha} \int_\bk F_{a_z}(\bm{\xi},\bk),\\
a_\perp & = & -\frac{1}{2\alpha} \int_\bk F_{b_\perp}(\bm{\xi},\bk),\\
b_\perp & = & -\frac{1}{2\alpha} \int_\bk F_{a_\perp}(\bm{\xi},\bk),\\
m  & = &  \frac12 \frac{1}{1+\delta+3\alpha} \int_\bk F_{m}(\bm{\xi},\bk),
\end{eqnarray}
where we have defined
\begin{equation}
F_\xi (\bm{\xi},\bk) = \sum_{n=1,2} \tanh\left(   \frac{|\epsilon_n(\bm{\xi}, \bk)|}{2 t}   \right)\,\partial_\xi |\epsilon_n(\bm{\xi}, \bk)|.
\label{eq.Ffunc}
\end{equation}

\section{Kitaev Majorana Fermions}
\label{sec.Kitaev}

We will now discuss the mean-field scheme based on the local mapping of the spin-1/2 operators $\hat \sigma^\gamma_i$ ($\gamma=x,y,z$) to 
a set of four Majorana fermion operators $\hat \eta^\mu_i$, ($\mu=0,x,y,z$) on each lattice site $i$, as discussed in the seminal paper by Kitaev \cite{Kitaev06}.
The Majorana fermion operators satisfy $(\hat \eta_i^\mu)^\dagger = \hat \eta_i^\mu$ and the Clifford algebra
\begin{equation}
\label{eq.Clifford}
\{\hat \eta_i^\mu,\hat \eta_j^\nu   \} = 2 \delta_{ij}\delta_{\mu\nu}.
\end{equation}
In terms of the Majorana fermions the spin operators are expressed as
\begin{equation}
\label{eq.KitaevMajorana}
\hat \sigma^\gamma_i = i \hat\eta^0_i \hat \eta^\gamma_i.
\end{equation} 

This Majorana representation of spins is over-complete and the physical Hilbert space is obtained by imposing the local constraint
$\hat \eta^0_i\hat \eta^x_i\hat \eta^y_i\hat \eta^z_i = 1$. It is indeed straightforward to check that the constraint ensures that the 
spin commutator relations are preserved. Using the properties of the Majorana fermion operators, Eq.~(\ref{eq.Clifford}),  it is possible to 
rewrite the constraint as \cite{Yilmaz+22}
\begin{equation}
\hat\eta^0_i \hat \eta^\gamma_i +\frac12 \epsilon_{\alpha\beta\gamma}\hat\eta^\alpha_i \hat \eta^\beta_i = 0,
\end{equation} 
which is quadratic in the Majorana operators. Since the antiferromagnetism will develop along the $z$ direction in spin space, it is sufficient 
to use the constraint $\hat\eta^0_i \hat \eta^z_i + \hat\eta^x_i \hat \eta^y_i=0$. We follow Ref. \cite{Yilmaz+22} and impose the constraint through a 
Lagrange multiplier field $\lambda_i$, 
\begin{equation}
\delta \hat {\cal H}_\lambda = i K\sum_i \lambda_i ( \hat\eta^0_i \hat \eta^z_i + \hat\eta^x_i \hat \eta^y_i).
\end{equation}

While $\lambda_i$ is not expected to enlarge the two-site unit cell of the honeycomb lattice, it could take different values on the $A$ and $B$ 
sites within the unit cell. We found that $\lambda_A = \lambda_B=0$ in the gapless and gapped QSL phases and $\lambda_A = -\lambda_B\neq 0$
in the antiferromagnetic phase. From now on we will therefore only include a single Lagrange multiplier 
\begin{equation}
\lambda=\lambda_A = -\lambda_B.
\end{equation}

In the following we will define $\hat c_i = \hat \eta_i^0$ to aid comparison with the mean-field theory based on the two-dimensional JWT. 
Expressing the spin operators in terms of Majorana fermions, using Eq.~(\ref{eq.KitaevMajorana}), the Hamiltonian (\ref{eq.Ham_spin}) 
contains only quartic interaction terms, 
\begin{eqnarray}
\hat {\cal H}/K & = & \sum_\br \left\{ \hat c_A(\br)\hat c_B(\br+\ba_1)\hat\eta_A^x(\br)\hat\eta_B^x(\br+\ba_1)\right.\\
& & + \hat c_A(\br)\hat c_B(\br+\ba_2)\hat\eta_A^y(\br)\hat\eta_B^y(\br+\ba_2)\nn\\
& & +\left.(1+\delta+\alpha) \hat c_A(\br)\hat c_B(\br)\hat\eta_A^z(\br)\hat\eta_B^z(\br)\right\}\nn\\
& & +\alpha \sum_\br \sum_{i=1,2}  \hat c_A(\br)\hat c_B(\br+\ba_i)\hat\eta_A^z(\br)\hat\eta_B^z(\br+\ba_i).\quad\nn
\end{eqnarray}

\begin{figure}[t]
 \includegraphics[width=0.6\linewidth]{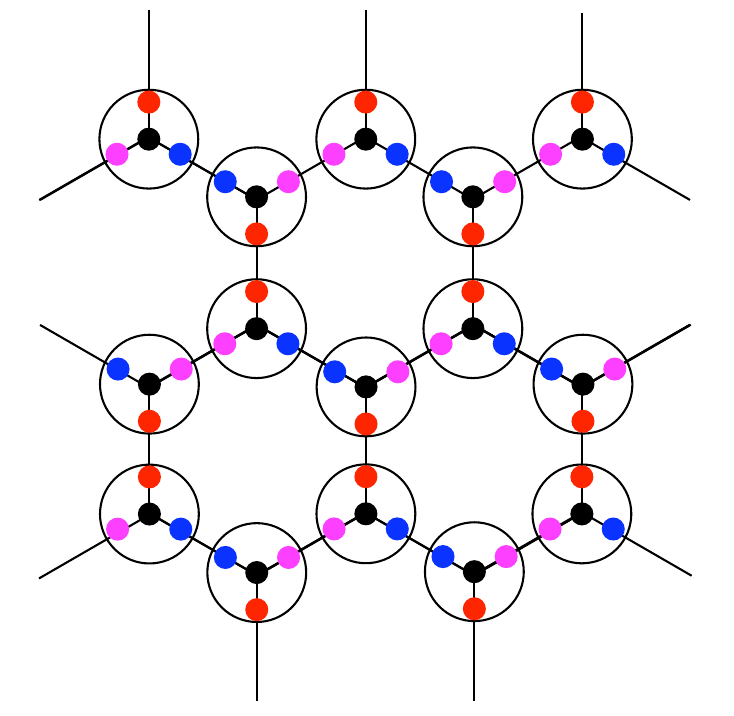}
 \caption{Using the Kitaev construction, each $S=1/2$ spin operator is represented in terms of a set of four Majorana fermions, subject to a local constraint.}
\label{figure3}
\end{figure}

\subsection{Mean-field theory}
As for the case of the 2$d$ JWT we will perform a simultaneous mean-field decoupling in the bond and site-diagonal magnetic channels. We introduce
the bond mean-field parameters  
\begin{eqnarray}
A^\gamma_{\br,\br'} & = &i  \langle  \hat\eta_A^\gamma (\br) \hat\eta_B^\gamma (\br')  \rangle,\\
B_{\br,\br'} & = &i  \langle  \hat c_A (\br) \hat c_B (\br')  \rangle,
\end{eqnarray}
 and define $A_\perp = A^x_{\br,\br+\ba_1} = A^y_{\br,\br+\ba_2}$, $A_\perp' =  A^z_{\br,\br+\ba_i}$, 
 $A_z =  A^z_{\br,\br}$, $B_\perp =  B_{\br,\br+\ba_i}$, $B_z = B_{\br,\br}$
for the relevant nearest neighbor bonds. The staggered magnetization is given by the expectation 
values
\begin{equation}
m = i \langle \hat c_A(\br) \hat\eta_A^z(\br)\rangle = - i \langle \hat c_B(\br) \hat\eta_B^z(\br)\rangle.
\end{equation}

After mean-field decoupling and Fourier transformation, as defined in Eqs. (\ref{eq.FT1},\ref{eq.FT2}), 
the mean-field Hamiltonian is 
\begin{align} 
&\frac{\hat{\cal H}_\textrm{mf}+\delta{\cal H}_\lambda}{KN} =\frac{i}{N} \sum_\bk \Big\{\nn\\
& \hat\Psi^\dagger_\bk\left(\begin{array}{cccc}0 & -\gamma_c^*& - (M-\lambda) & 0 \\ \gamma_c& 0 & 0 &  (M-\lambda) \\  (M-\lambda) & 0 & 0 & -\gamma^*_z \\ 0 & - (M-\lambda) & \gamma_z & 0  \end{array} \right)
\hat \Psi_\bk\nn\\
& + \hat\Phi^\dagger_\bk\left(\begin{array}{cccc}0 & -\gamma_x^*&  \lambda & 0 \\ \gamma_x& 0 & 0 & - \lambda \\ - \lambda & 0 & 0 & -\gamma^*_y \\ 0 & \lambda & \gamma_y & 0  \end{array} \right)
\hat \Phi_\bk\Big\}\nn\\
& + (1+\delta+\alpha)A_z B_z +2 \tilde{A}_\perp B_\perp+(1+\delta+3\alpha) m^2,
\end{align}
where $\Psi_\bk = \left(\hat c_A(\bk), \hat c_B(\bk), \hat\eta_A^z(\bk), \hat\eta_B^z(\bk)\right)^T$, $\Phi_\bk = \left(\hat\eta_A^x(\bk), \hat\eta_B^x(\bk), \hat\eta_A^y(\bk), \hat\eta_B^y(\bk)\right)^T$, $M=(1+\delta+3\alpha)m$, $\tilde{A}_\perp = A_\perp+\alpha A_\perp'$, 
\begin{eqnarray}
\label{eq.gammac2}
\gamma_c (\bk) & = &  (1+\delta+\alpha) A_z + \tilde{A}_\perp \left( e^{i\bk\ba_1}+e^{i\bk\ba_2}  \right),\\
\gamma_z (\bk) & = &  (1+\delta+\alpha) B_z + \alpha B_\perp \left( e^{i\bk\ba_1}+e^{i\bk\ba_2}  \right),
\end{eqnarray}
$\gamma_x(\bk)  =  B_\perp  e^{i\bk\ba_1}$ and $\gamma_y(\bk)  =  B_\perp  e^{i\bk\ba_2}$. The resulting energy eigenvalues $\pm|\epsilon_{1,2}(\bk)|$ and $\pm|\epsilon_{3,4}(\bk)|$ are given by 
\begin{eqnarray}
\label{eq_dispKitaev1}
\epsilon_{1,2}^2(\bk) & = &  \frac{|\gamma_c|^2+|\gamma_z|^2}{2}+(M-\lambda)^2\\
& & \pm \sqrt{\left( \frac{|\gamma_c|^2-|\gamma_z|^2}{2}  \right)^2+|\gamma_c-\gamma_z|^2  (M-\lambda)^2},\nn\\
\label{eq_dispKitaev2}
\epsilon_{3,4}^2(\bk) & = & B_\perp^2+\lambda^2 \pm \lambda| \gamma_x-\gamma_y      |\\
& = &  B_\perp^2+\lambda^2 \pm 2 B_\perp \lambda \sin\left(\frac{\sqrt{3}}{2}k_x   \right).\nn
\end{eqnarray}

\noindent
Minimizing the free-energy density,
\begin{eqnarray}
\label{eq.en2}
f &  = &   -t \sum_{n=1}^4\sum_{\sigma=\pm1}\int_\bk \ln\left(e^{-\sigma|\epsilon_n(\bk)|/t}+1  \right)\\
& & + (1+\delta+\alpha)A_z B_z +2 \tilde{A}_\perp B_\perp+(1+\delta+3\alpha)m^2,\nn
\end{eqnarray}
with respect to $\xi\in\{A_z, B_z,  \tilde{A}_\perp, B_\perp, m \}$, $\partial_\xi f =0$,  
we obtain the self-consistency equations
\begin{eqnarray}
A_z  & = &   \frac{1}{1+\delta+\alpha} \int_\bk F_{B_z}(\bm{\xi},\bk),\\
B_z  & = &   \frac{1}{1+\delta+\alpha} \int_\bk F_{A_z}(\bm{\xi},\bk),\\
\tilde{A}_\perp & = & \frac12 \int_\bk F_{B_\perp}(\bm{\xi},\bk),\\
B_\perp & = & \frac12 \int_\bk F_{\tilde{A}_\perp}(\bm{\xi},\bk),\\
m  & = &  \frac12 \frac{1}{1+\delta+3\alpha} \int_\bk F_{m}(\bm{\xi},\bk),
\end{eqnarray}
where the functions $F_\xi(\bm{\xi},\bk)$ are defined as in Eq.~(\ref{eq.Ffunc}) but with the 
sum running over the four bands $n=1,\ldots,4$ given in Eqs. (\ref{eq_dispKitaev1}) and (\ref{eq_dispKitaev2}).

\subsection{Determination of the Lagrange multiplier}
\label{sec.Kitaev_lagrange}

\begin{figure}[t]
 \includegraphics[width=\linewidth]{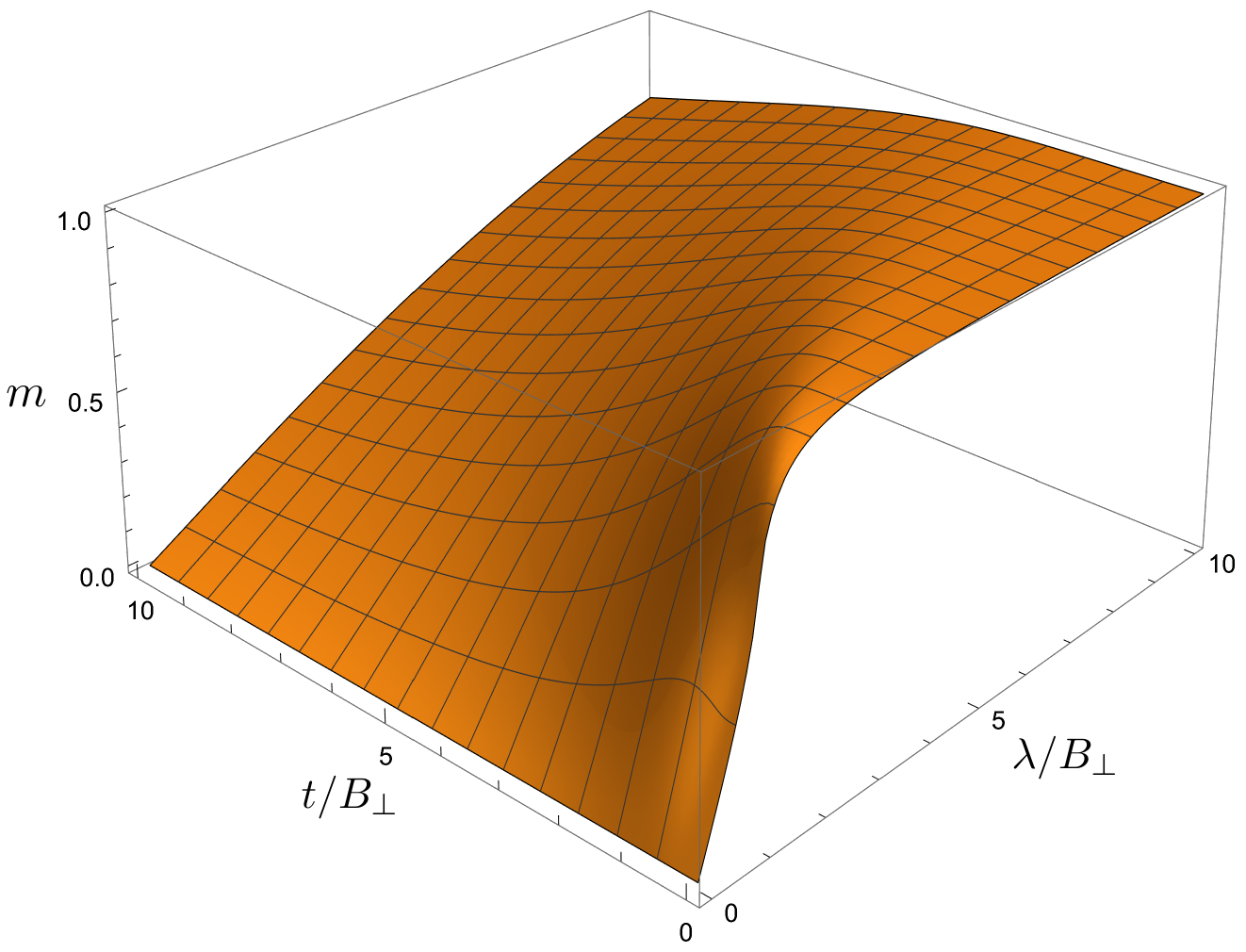}
 \caption{The function $m=\Omega(\lambda/B_\perp, t/B_\perp)$ relating the Lagrange multiplier $\lambda$ to the staggered magnetization $m$, the 
 bond mean-field parameter $B_\perp$ and the dimensionless temperature $t=T/K$.}
\label{figure4}
\end{figure}

The Lagrange multiplier $\lambda$ is closely linked to the staggered magnetization $m$, which satisfies the self-consistency equation
\begin{equation}
m = \frac{1}{2} \sum_{n=1,2}\int_\bk \tanh\left( \frac{|\epsilon_n(\bk)|}{2 t}  \right)\partial_M |\epsilon_n(\bk)|,
\end{equation}
where $M=(1+\delta+3\alpha)m$. Note that the bands $n=3,4$ do not depend on $M$. From $\partial_\lambda f = 0$ and 
using that $\partial_\lambda\left\{ |\epsilon_1(\bk)|+|\epsilon_2(\bk)|\right\} = -\partial_M \left\{|\epsilon_1(\bk)|+|\epsilon_2(\bk)|\right\}$ 
we obtain
\begin {eqnarray}
m  & = &  \frac12 \sum_{n=3,4}\int_\bk  \tanh\left( \frac{|\epsilon_n(\bk)|}{2 t}  \right) \partial_\lambda |\epsilon_n(\bk)|\nn\\ 
& = &  \Omega\left(\frac{\lambda}{B_\perp},\frac{t}{B_\perp}\right),
\end{eqnarray}
where the function $\Omega$ is independent of $m$ and given by the momentum integral
\begin{eqnarray}
\Omega(x,y) & = &  \frac{1}{2} \sum_{\kappa=\pm1}\int_\bk \partial_x \sqrt{1+x^2+2\kappa x\sin\left(\sqrt{3}/2\,k_x   \right)}\nn\\
& & \times \tanh\left(\frac{\sqrt{1+x^2+2\kappa x\sin\left(\sqrt{3}/2\,k_x   \right)} }{2y}   \right).
\end{eqnarray}

It is straightforward to compute the function $\Omega(x,y)$ numerically. The resulting relation between $m$, $\lambda/B_\perp$ and $t/B_\perp$ is shown in Fig.~\ref{figure4} 
for positive values of the staggered magnetization. The domain of negative magnetizations is obtained for negative Lagrange multipliers, $\Omega(-x,y)=-\Omega(x,y)$.
This result shows that the Lagrange multiplier is zero in the non-magnetic  phases and non-zero if the staggered magnetization is finite. 

In the following, we will minimize the free-energy density $f$ (\ref{eq.en2}) at given temperature $t$ with respect to the mean-field parameters $A_z$, $B_z$, $\tilde{A}_\perp$,  $B_\perp$ and  $m$ by solving the 
corresponding self-consistency integral equations iteratively. At each step of the iteration we determine the Lagrange multiplier $\lambda$ from the values 
of $B_\perp$ and $m$, using the equation $m=  \Omega(\lambda/B_\perp, t/B_\perp)$.

\section{Results}
\label{sec.results}

\subsection{Topological Phase Transition}
\label{sec.results_top}

We start by discussing the zero-temperature topological phase transition between the gapless and gapped Kitaev QSL states as a function of 
the anisotropy $\delta=(K_z-K)/K$ of the Kitaev couplings and the relative strength $\alpha=J/K$ of the Ising coupling. 

For $\alpha=0$ the Kitaev model is exactly solvable and the topological phase transition is known to occur at $\delta_c=1$, as derived in the triangle 
inequalities in Kitaev's original paper \cite{Kitaev06}. 
At this point the Dirac points of the dispersive low energy band  merge, forming a semi Dirac point. We expect that the anisotropy 
induced by the Ising coupling $\alpha$ has a similar effect. However, sufficiently strong $\alpha$ will induce an antiferromagnetic state, 
which we will consider later. 

In the absence of magnetization, $m=0$,  the dispersion of the low-energy band is simply given by $\pm |\gamma_c(\bk)|$, where 
$\gamma_c(\bk) = u+v\left(e^{i\bk\ba_1}+e^{i\bk\ba_2} \right)$. The coefficients $u$ and $v$ are functions of $\delta$, $\alpha$ and of 
the mean-field parameters, as defined in Eq.~(\ref{eq_gammac}) for the JWT and in Eq.~(\ref{eq.gammac2}) for the Kitaev representation. 

Rather than computing the momentum separation of the Dirac points in the gapless QSL or the the size of the energy gap on the other side of the 
transition,  it is more convenient to compute the ratio $r=|u/v|$. While for $r<2$ the band exhibits gapless Dirac points at zero energy, for 
$r>2$ the Majorana fermion spectrum becomes gapped. We can therefore simply determine the topological phase transition at $r_c=2$ by using 
a bi-section method. 

Let us first determine the topological phase boundary using the mean-field theory based on the JWT. In this case the parameter $r=|u/v|$ is given by 
\begin{equation}
r:=\frac{(1+\delta+\alpha) |a_z|}{1+\alpha a_\perp}.
\end{equation} 

 For $\alpha=0$ the Kitaev model is exactly solvable 
in terms of Jordan-Wigner fermions since the local operator $(\hat c^\dagger_{A,\br}+\hat c_{A,\br}) (\hat c^\dagger_{B,\br}-\hat c_{B,\br})= -i \hat \eta_A^z (\br)\hat \eta_B^z(\br)$ 
commutes with the Hamiltonian. Although we don't require  a mean-field treatment in this case, it is interesting to understand how the correct value of the topological 
phase transition, $\delta_c =1$ ($K_z/K=2$),  is recovered within our mean-field theory. In fact, the mean-field theory becomes trivial for the pure anisotropic Kitaev model since the 
local flux excitations are dispersionless, with corresponding bands at energies $\pm|\gamma_z(\bk)| = \pm (1+\delta)|b_z|$. The free energy is independent of the mean-field parameters 
$a_\perp$ and $b_\perp$ and at zero temperature, $t=0$,  we obtain the mean-field 
parameters $a_z$ and $b_z$ from minimizing the energy 
\begin{eqnarray}
\varepsilon(a_z,b_z) & = &  -\frac{1}{V_{\cal{BZ}}} \int_{\cal{BZ}} d^2\bk \left| (1+\delta)a_z  +e^{i\bk\ba_1}+e^{i\bk\ba_2}   \right| \nn\\
& & -(1+\delta)(|b_z|+a_z b_z).
\end{eqnarray}

\begin{figure}[t]
 \includegraphics[width=\linewidth]{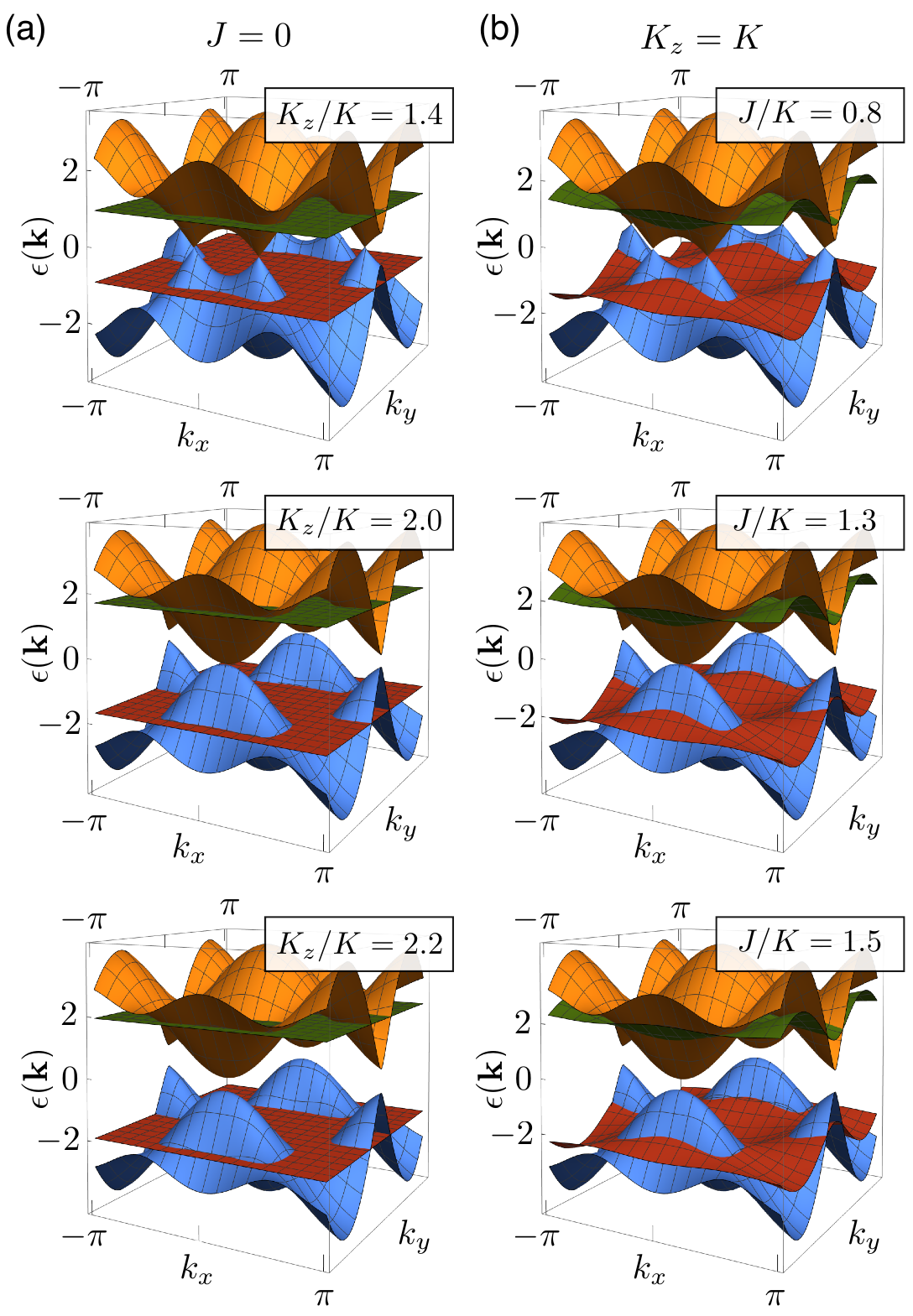}
 \caption{Evaluation of the Majorana fermion mean-field spectra across the topological phase transition between a gapless and gapped Kitaev QSL, 
 as a function of (a) the anisotropy $K_z/K$ in the pure Kitaev model and (b) the Ising coupling $J/K$ in the case $K_z=K$. Note that the 
 two mean-field schemes give identical results.  At the topological phase transition the Dirac points merge along the edge of the Brillouin zone, as schematically 
 shown in Fig.~\ref{figure1}(b).}
\label{figure5}
\end{figure}

From $\partial \varepsilon/\partial b_z = 0$ we obtain $a_z=1$ if $b_z<0$ and $a_z=-1$ if $b_z>0$, regardless of the value of the anisotropy $\delta$. Let us focus on the first case. Inserting $a_z=1$ into  
$\partial \varepsilon/\partial a_z = 0$ we obtain
\begin{equation}
b_z =  -\frac{1}{V_{\cal{BZ}}} \int_{\cal{BZ}} d^2\bk \frac{\cos(\bk\ba_1)+\cos(\bk\ba_2)}{ \left| 1+\delta  +e^{i\bk\ba_1}+e^{i\bk\ba_2}   \right|},
\end{equation}
which is indeed negative. Note that for $a_z=-1$ we obtain the same value for $b_z$ but with positive sign. This solution is equivalent to the first solution but 
with a momentum shift of the entire excitation spectrum. We obtain $r=1+\delta$, and hence a topological phase transition at $\delta_c=1$, 
which is equivalent to $K_z/K =2$. 

\begin{figure}[t]
 \includegraphics[width=0.85\linewidth]{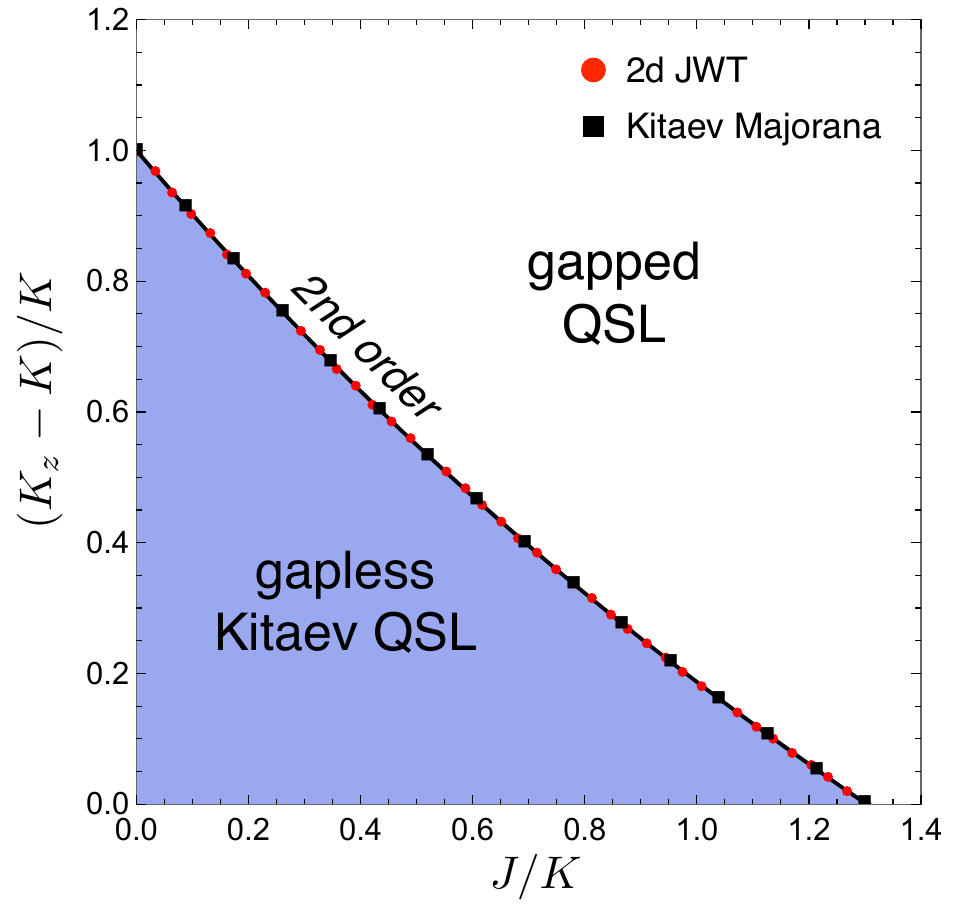}
 \caption{Topological phase boundary between a gapless and gapped Kitaev QSL as a function of the Ising coupling $\alpha=J/K$ and the anisotropy $\delta=(K_z-K)/K$ of the 
 Kitaev couplings. Note that the mean field schemes based on the two-dimensional JWT and on the Kitaev Majorana representation give identical results. Potential magnetic instabilities
 are not considered here.}
\label{figure6}
\end{figure}

The resulting mean-field spectra are shown in Fig.~\ref{figure5}(a) for different values of $\delta$. With increasing $\delta$ the Dirac points approach each other along one of the edges of the hexagonal 
Brillouin zone, as illustrated in Fig.~\ref{figure1}(b), and merge at $\delta_c=1$, forming a semi-Dirac point. For $\delta>1$ the spectrum becomes gapped. The other bands remain gapped and dispersionless 
across the topological phase transitions and only slightly change in energy.    

As a next step we investigate the effect of the Ising coupling $\alpha=J/K$ on the isotropic Kitaev model, $\delta=0$ ($K_z=K$). From Eqs. (\ref{eq_gammac}) and (\ref{eq_gammaeta}) it is clear that the gapped bands of flux excitations become 
weakly dispersive and that the Ising coupling induces anisotropy in the gapless Majorana bands. For $\alpha>0$ the mean-field theory is no longer trivial and the free energy becomes a function of the four mean-field 
parameters  $a_z$, $b_z$, $a_\perp$ and $b_\perp$. As shown in Fig.~\ref{figure5}(b), the effect of the Ising coupling on the gapless Dirac band is very similar to that of the anisotropy in the Kitaev couplings, and the system undergoes a topological 
phase transition at $\alpha_c\approx 1.3$. However, we expect that this transition will be pre-empted by an antiferromagnetic instability. The topological phase boundary as a function of both, $\delta$ and $\alpha$ is shown in Fig.~\ref{figure6}. 

We briefly discuss the mean-field theory for the topological phase transition base on the Kitaev Majorana representation. This treatment involves a larger number of degrees of freedom with one gapless Dirac band and three gapped bands
of flux excitations. In addition, we have to incorporate a Lagrange multiplier $\lambda$ to enforce the local constraints on the Majorana fermions. However, as we have seen in Sec.~\ref{sec.Kitaev_lagrange}, $\lambda$ is identical to zero 
in the non-magnetic QSL states. This leads to a drastic simplification of the spectrum since for $\lambda=0$ the $\eta_x$ and $\eta_y$ bands don't hybridize and remain flat across the topological transition
with degenerate energies $\pm|\gamma_x(\bk)|=\pm|\gamma_y(\bk)|=\pm B_\perp$, even if the Ising coupling $\alpha$ is included. For all values of the anisotropy $\delta$ and the Ising coupling $\alpha$ 
the mean-field dispersions $\pm|\gamma_c(\bk)|$ and $\pm|\gamma_z(\bk)|$ of the other two Majorana bands  
are identical to those of the two Majorana bands in the JWT treatment. It is therefore not surprising that the two mean-field treatments result in identical phase boundaries for the topological phase transition between the gapless 
and gapped Kitaev QSLs, as shown in Fig.~\ref{figure6}.

\begin{figure}[t]
 \includegraphics[width=0.87\linewidth]{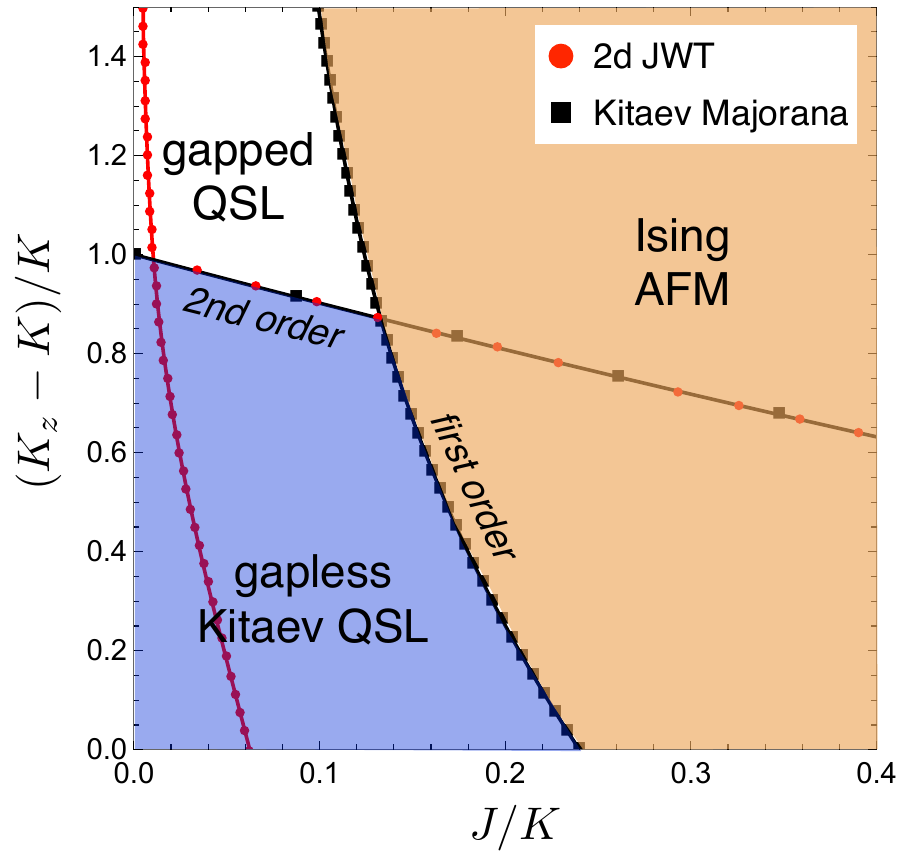}
 \caption{Zero-temperature phase diagram of the anisotropic Kitaev-Ising model as a function of the relative strength $\alpha=J/K$ of the Ising coupling and the anisotropy 
 $\delta=(K_z-K)/K$ of the Kitaev couplings. While the two mean-field theories give identical phase boundaries for the topological phase transition between the gapless and gapped Kitaev QSLs, 
 the treatment based on the JWT fails to correctly describe the first-order transition to the antiferromagnetically ordered state.}
\label{figure7}
\end{figure}

\subsection{Antiferromagnetism}

In the previous section we have not included the possibility of the formation of an antiferromagnetic state with finite staggered magnetization $m$. For $m\neq 0$ the Lagrange multiplier $\lambda$ in the Kitaev Majorana mean-field theory 
is no longer zero and acquires a value of the order of the magnetization (see Fig.~\ref{figure4}). As a result, the $\eta_x$ and $\eta_y$ Majorana fermions hybridize and form dispersive bands with energies 
$\pm \epsilon_{3,4}(\bk)$ (\ref{eq_dispKitaev2}). This shows that the $\eta_x$, $\eta_y$ fermions are not simply spectators as in the case of the topological phase transition but play a crucial role in the energetics of the antiferromagnetic 
transition. Since these degrees of freedom are neglected in the JWT with fixed string orientation, we expect that the corresponding mean-field theory does not correctly describe the magnetic instability. 

As shown in Fig.~\ref{figure7}, there is indeed a significant discrepancy between the zero-temperature magnetic phase boundaries calculated within the two mean-field theories.  We find that the magnetic phase transition is strongly first order 
with a jump in magnetization close to the fully polarized value.  This is not surprising. The antiferromagnetic ordering is driven by an Ising exchange and as a result quantum fluctuations are frozen out at low temperatures, resulting in a 
large ordered moment. Moreover, transverse spin fluctuations are active only along the one-dimensional zig-zag chains formed by the $x$ and $y$ bonds.

\subsection{Finite Temperature Phase Diagram}

In the previous section we have identified problems with the mean-field theory based on the JWT for states with finite magnetization. Neglecting the flux excitations of the $\hat\eta^x$ and $\hat\eta^y$ 
Majorana fermions by using a particular gauge choice of the string operator in the 2d JWT, we also induce pathologies at finite temperatures. In order to illustrate this we focus on the pure isotropic Kitaev 
model. In this case we obtain the mean-field dispersions $\epsilon_1(\bk) = |\gamma_c(\bk)| = |a_z+e^{i\bk\ba_1}+e^{i\bk\ba_2}|$ for the dispersive Dirac band and $\epsilon_2(\bk)=|\gamma_z(\bk)|=|b_z|$
for the flat band.  As discussed in Sec.~\ref{sec.results_top}, at zero temperature the minimization of the energy density $\varepsilon(a_z,b_z)$ with respect to $a_z$ and $b_z$ results in two equivalent 
mean-field solutions, e.g. one with a negative value of $b_z$ and $a_z=1$. This reproduces the correct excitation spectrum of the isotropic Kitaev model with Dirac points at the corners of the hexagonal Brillouin zone.   
Let us now investigate the mean-field solution at finite temperature $t=T/K$. Minimizing the free energy density $f(a_z,b_z)$ with respect to $b_z$ we obtain the simple relation
\begin{equation}
a_z = -\tanh\left(\frac{b_z}{2 t }   \right).
\end{equation}   
While for $b_z<0$ we recover $a_z=1$ as $t\to 0$, at any finite temperature $a_z<1$. This corresponds to a mean-field dispersion with Dirac points displaced along the edges of the Brillouin zone, in the same way as
for an anisotropic Kitaev model with $K_z<K$. This is clearly unphysical and caused by an artificial symmetry breaking due to the fixed orientation of the string operator.  

\begin{figure}[t]
 \includegraphics[width=\linewidth]{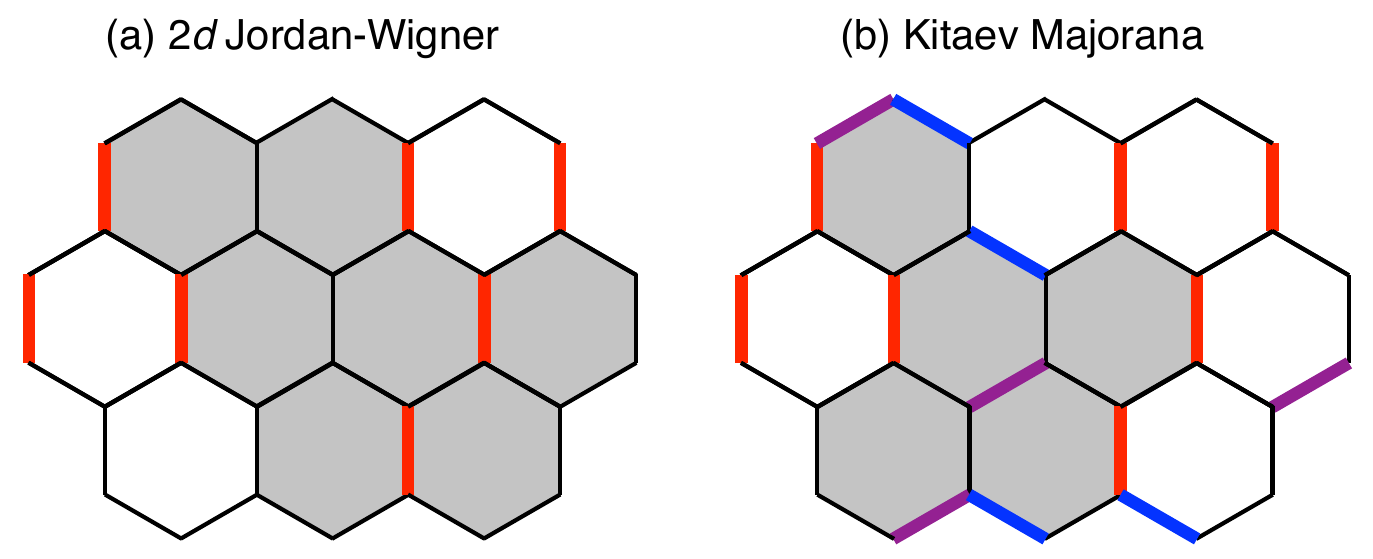}
 \caption{Illustration of the flux excitations in terms of the (a) 2$d$ JWT and (b) the Kitaev Majorana representation. Excited bonds are shown as thick lines and plaquettes 
 with non-zero flux are shaded grey.}
\label{figure8}
\end{figure}

The different finite-temperature response of the JWT and Kitaev-Majorana mean-field treatments can also be understood in terms of thermal excitations of the gapped flux excitations. 
As pointed out by Kitaev \cite{Kitaev06}, expressing the spin-1/2 operators as $\hat\sigma_i^\gamma= i\hat\eta_i^0\hat\eta_i^\gamma$  ($\gamma=x,y,z$), the $Z_2$ flux through a hexagon is 
given by the product $\hat W_p$ of the nearest-neighbor bond operators $\hat A_{\langle i, j\rangle_\gamma}^\gamma= i \hat\eta_i^\gamma\hat\eta_j^\gamma$ around the plaquette. The bond operators 
square up to the identity operator and hence have eigenvalues plus or minus one. The plaquette carries a flux if an odd number of bonds around it are excited ($\hat W_p = -1$). In the case of the pure 
(anisotropic) Kitaev model, the bond operators are local and $\hat W_p$ commutes with the Hamiltonian. As a result, the Hamiltonian can be diagonalised for each flux configuration and the ground state corresponds 
to the zero flux sector, resulting in a non-interacting  Hamiltonian for the dispersive $\hat\eta^0$ Majorana fermion. 

Both the 2$d$ JWT and the Kitaev Majorana approaches correctly describe the zero flux sector and hence the ground-state properties of the anisotropic Kitaev model. At first glance, it might seem that the 
two approaches enumerate flux excitations differently since the JWT only includes  bond excitations on the $z$ links. However, the choice of the string operator in the JWT is a gauge degree of freedom and  
the Kitaev Majorana fermions are subject to local constraints. In the end, both mappings are exact and therefore equivalent.  The problems arise when the finite-temperature mean-field average over bond 
operators is taken for a fixed orientation of the string.  In Fig.~\ref{figure8} the bond excitations and resulting fluxes are sketched for the two approaches. 

It is also worth mentioning that for the particular Kitaev-Ising model only the bond excitations on the $z$ links acquire dynamics, resulting in the same mean-field dispersion of the $\hat \eta_z$ Majorana fermion as
in the mean-field treatment based on the JWT. This is the reason why both approaches result in the same zero-temperature phase boundary for the topological phase transition between the gapless and gapped QSL states.  
Note that this is a special feature of the Ising exchange $J$. For a Heisenberg coupling the $\hat \eta_x$ and $\hat \eta_y$ fermions acquire dynamics as well.

\begin{figure}[t]
 \includegraphics[width=0.95\linewidth]{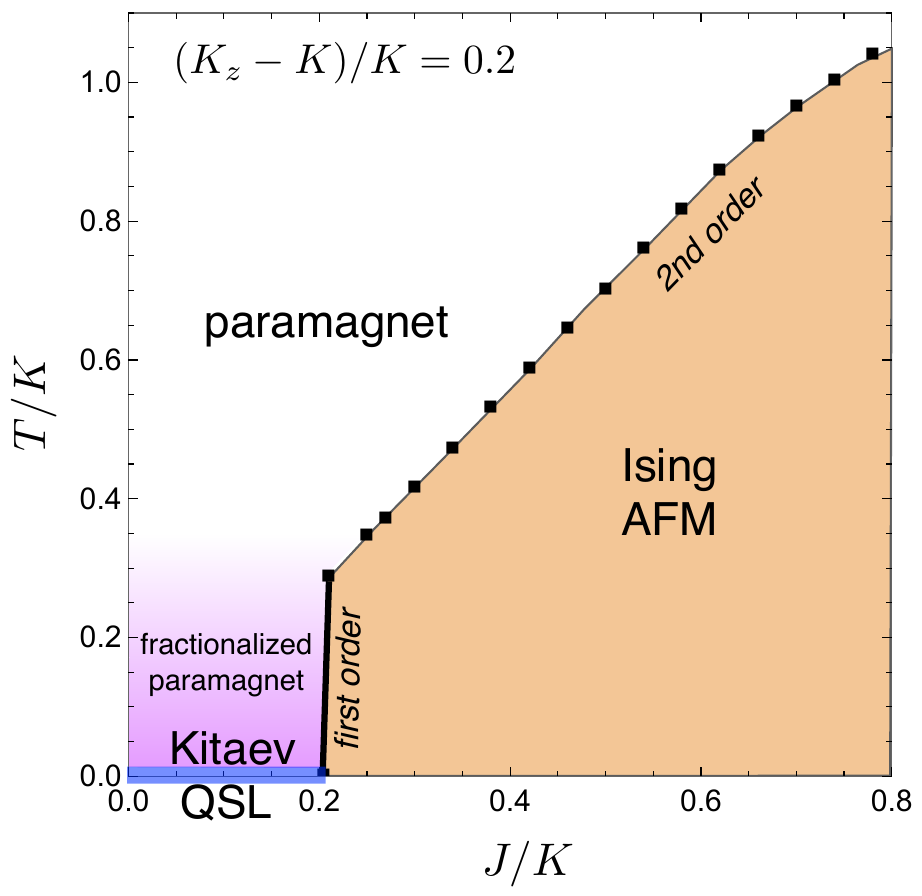}
 \caption{Finite-temperature mean-field phase diagram of the anisotropic Kitaev-Ising model as a function of the relative strength of the Ising coupling $\alpha=J/K$
 and the dimensionless temperature $t=T/K$  for a value of $\delta=0.2$ of the anisotropy of the Kitaev couplings.}
\label{figure9}
\end{figure}

Because of the problems with the finite temperature mean-field theory  based on the two-dimensional JWT, we will use the  
Kitaev Majorana fermion representation to determine finite temperature phase diagrams,
following the procedure outlined in Sec.~\ref{sec.Kitaev}. In Fig.~\ref{figure8} a representative phase diagram is shown as a function of the Ising coupling $\alpha=J/K$ and the dimensionless temperature $t=T/K$ for 
a fixed values $\delta=0.2$ of the anisotropy of the Kitaev couplings. For this value of $\delta$ we find a zero-temperature phase transition from a gapless Kitaev QSL to an antiferromagnetic state at $\alpha_c\approx 0.2$.
This transition is strongly first order. At small temperatures the antiferromagnetic transition remains first order and is very steep.  As one might expect, the magnetic transition becomes 
continuous above a certain temperature.

Let us now investigate the finite-temperature behavior in the regime of small values of $J/K$ where the zero-temperature ground state is a gapless Kitaev QSL. An important energy scale is the gap $\Delta$ of the 
flux excitations which is equal to $\Delta/K\approx 0.26$ for the isotropic Kitaev model \cite{Kitaev06}. While for $T\ll\Delta$ the typical separation between fluxes is exponentially large and the thermal average of the flux operator 
$\langle \hat W_p\rangle$ close to $+1$, at temperatures 
$T>\Delta$, the flux excitations proliferate with high probability on all plaquettes, resulting in $\langle \hat W_p\rangle = 0$ of the flux operator.  One might therefore expect a finite-temperature confinement transition  
from a QSL with deconfined Majorana fermions to a paramagnet where the Majorana fermions are confined  
 via the flux excitations of the emergent $Z_2$ gauge field \cite{hermanns+18}. However, it is known that in two dimensions gauge theories are confining at any non-zero temperature. Hence the Kitaev QSL exists only at zero temperature
 and even an exponentially small density of thermally excited fluxes is sufficient to destroy the QSL state.  Nevertheless, interesting finite temperature crossovers are seen in quantum Monte Carlo simulations of the two-dimensional Kitaev model, 
 using Majorana fermion representations \cite{Nasu+14,Nasu+15,Motome+20}. 
 
\begin{figure}[t]
 \includegraphics[width=0.9\linewidth]{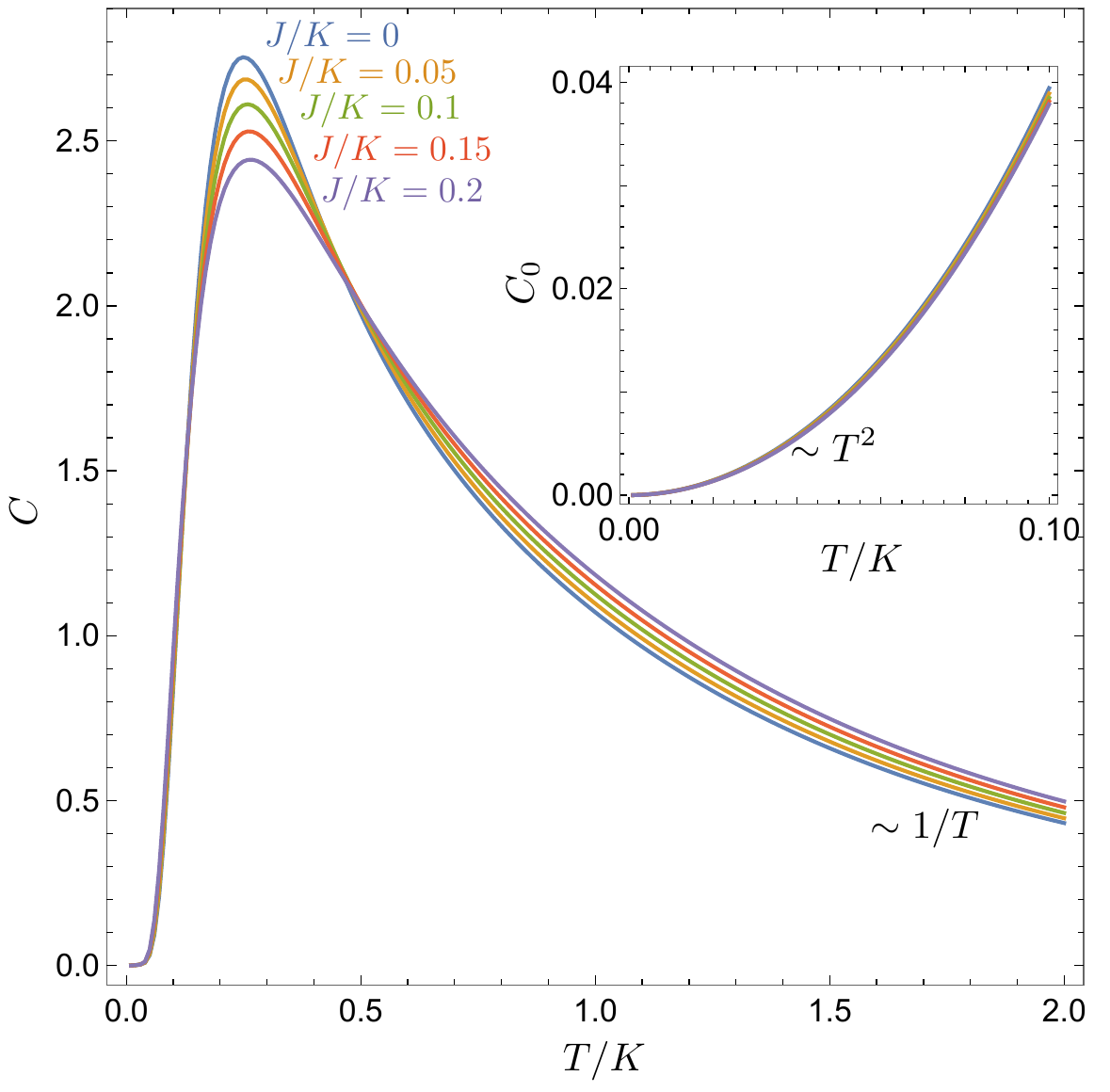}
 \caption{Specific heat $C$ per unit cell as a function of the dimensionless temperature $T/K$, for an anisotropy $\delta=0.2$ of the Kitaev couplings and different values of 
 $\alpha=J/K$, corresponding to a gapless Kitaev QSL ground state. The peak is located at the energy scale of the gapped flux excitations and indicates a crossover from 
 a fractionalized paramagnet with frozen flux excitations to a conventional paramagnet at high temperatures. In the latter the expected Curie dependence $C\sim 1/T$ is observed. The inset shows the 
 low-temperature specific heat contribution $C_0$ of the gapless Majorana fermion band, showing the $T^2$ dependence expected for Dirac fermions in $d=2$.}
\label{figure10}
\end{figure}

 In order to identify finite-temperature crossovers, we compute the specific heat per unit cell, 
 \begin{equation}
 C = -t\frac{\partial^2 f}{\partial t^2},
 \end{equation} 
 where $f(t)$ denotes the mean-field free energy density $f=F/(N K)$ (\ref{eq.en2}) as a function of the dimensionless temperature $t=T/K$. In Fig.~\ref{figure10} the temperature evolution of the specific heat is 
 shown for systems with an anisotropy $\delta=0.2$ of the Kitaev couplings and increasing values $\alpha=J/K$ of the Ising coupling, up to the value $\alpha=0.2$, which is slightly below the critical value of the $t=0$ first-order 
 transition between the Kitaev QSL and the Ising antiferromagnet.  Unlike in previous work using quantum Monte Carlo simulations of finite systems \cite{Nasu+14,Nasu+15,Motome+20}, where 
 two separate specific-heat peaks are found,  our mean-field results only show a single peak at a temperature $T\approx\Delta$, where $\Delta$ is the energy gap of flux excitations. Note that for the pure Kitaev model ($\alpha=0$) the 
 anisotropy $\delta=0.2$ gives rise to a small splitting of the flux gaps, $\Delta_z/K\approx 0.29$ and $\Delta_\perp/K\approx 0.24$, resulting in a slight broadening of the crossover peak in the specific heat. With increasing Ising 
 coupling $\alpha$ the splitting further increases up to values $\Delta_z/K\approx 0.32$ and $\Delta_\perp/K\approx 0.22$ for $\alpha=0.2$. Note that $\alpha$ also gives dynamics to the bond excitations along the $z$ links, 
 adding to the broadening of the crossover. At high temperatures, $T>\Delta$, we recover a conventional paramagnet, and the Curie-Weiss dependence $C\sim1/T$ is clearly observed above temperatures of the order of the bandwidth of the 
 Majorana fermions. At temperature $T<\Delta$ flux excitations are exponentially suppressed and signatures of fractionalization become visible. The crossover to a fractionalized paramagnet at low temperatures is indicated by 
 a color gradient in the phase diagram, Fig.~\ref{figure9}. The energy scale of the crossover coincides with the point at which the magnetic phase transition becomes continuous. 
 
 The inset of Fig.~\ref{figure10} shows the specific heat contribution from the gapless Majorana fermion band at lowest temperatures. As expected, we observe the $C_0\sim T^2$ dependence
 expected for Dirac fermions in two spatial dimensions.

\section{Discussion and Conclusion}
\label{sec.discussion}

In this paper we have determined zero- and finite-temperature phase diagrams of the anisotropic, antiferromagnetic Kitaev-Heisenberg model on the honeycomb lattice, using parton mean-field theories based on two 
different Majorana fermion representations of the $S=1/2$ spin operators: the one used by Kitaev \cite{Kitaev06} and a two-dimensional Jordan-Wigner transformation (JWT) \cite{Feng+07,Chen+07,Chen+08,Dora+18}.
Both mappings have been used to obtain the exact solution of the anisotropic Kitaev model \cite{Kitaev06,Chen+07}. 

In order to ensure that  the Hamiltonian remains local after JWT, we studied a particular limit of the model, keeping the anisotropy in the Kitaev couplings finite while taking the extreme limit of an infinitely strong 
Ising anisotropy in the Heisenberg sector. For this model, it is possible to use the same snake-string operators in the JWT as for the pure anisotropic Kitaev model \cite{Feng+07,Chen+07,Chen+08,Dora+18},  resulting 
in two Majorana modes. For sufficiently weak anisotropy and Ising coupling the low-energy band is gapless with Dirac points on the edges of the hexagonal Brillouin zone. The other band of gapped flux excitations is flat for 
the anisotropic Kitaev model but becomes weakly dispersive in the presence of the additional Ising coupling.

On the other hand, following Kitaev's approach \cite{Kitaev06}, the spin-$1/2$ operators are mapped to a set of four Majorana fermions with three modes corresponding to gapped flux excitations. The Majorana fermion operators are subject 
to local constraints which we reformulated in a quadratic form and enforced through a Lagrange multiplier, following a previous study \cite{Yilmaz+22} of the magnetic field dependence of the pure Kitaev model. 

Perhaps surprisingly, the two mean-field theories result in identical zero-temperature phase boundaries for the topological phase  transition between the gapless and gapped Kitaev QSL states. The reason is that  for an Ising 
exchange $\alpha=J/K$ only the bond excitations along the $z$ links, which are accounted for in both approaches, acquire dynamics.  The two additional gapped modes 
in the Kitaev mean-field theory remain flat across the transition and don't contribute to the physics. The  mean-field dispersion of the remaining two bands is identical to the mean-field spectrum based on the JWT. The 
mean-field treatments give the correct value $K_z/K=2$ for the topological phase transition of the anisotropic Kitaev model. The Ising coupling $J/K$ is an additional source of anisotropy and cooperates with the anisotropy 
in the Kitaev couplings in driving the topological phase transition.

We demonstrated that all three bands of flux excitations play a crucial role for the antiferromagnetic instability and the finite-temperature behavior. The mean-field theory based on the two-dimensional JWT 
therefore fails to correctly describe the finite temperature phase diagrams.  Even for an isotropic Kitaev model we found an anisotropic response at finite temperatures. 
We believe that this unphysical behavior is not an intrinsic problem of the two-dimensional JWT since the choice of the string operator is a gauge degree of freedom. However, taking a finite-temperature mean-field average 
over the bond operators leads to artificial anisotropy that depends upon the choice of the string. Recent progress has been made in formulating a JWT  in two and three dimensions that keeps locality and all relevant symmetries 
manifest \cite{Po21,Li+22}. This is achieved through operators that create local deformations 
of the JW string operator. A mean-field theory based on such a gauge invariant formulation of the JWT would not suffer from artificial symmetry breaking. Given the increased complexity it remains unclear, however, 
if this approach if useful for practical calculations when dealing with realistic spin Hamiltonians. 

We instead used the parton mean-field theory formulated in terms of the Kitaev Majorana fermions to obtain the finite-temperature phase diagram of the 
anisotropic Kitaev-Ising model. As expected, sufficiently strong Ising exchange results in a first-order transition from the gapless and gapped QSLs to an antiferromagnetic phase with fully gapped Majorana fermion spectrum. 
Unfortunately, we are not aware of numerical results in the literature for the magnetic instability of the antiferromagnetic Kitaev-Ising model. The critical mean-field value $(J/K)_c\approx 0.2$ for the first-order 
transition between the Kitaev QSL and the Ising antiferromagnet is considerably larger than the value $(J/K)_c\approx 0.035$ for the isotropic Kitaev-Heisenberg model, computed with tensor-network algorithms \cite{Osorio+14}. 

Although the QSL states 
only exist at zero temperature, the magnetic phase transition remains first order at low temperatures and becomes continuous  above a certain temperature.  While we believe that this 
behavior is generic and similar to other QSL systems, the first-order behavior is particularly strong for the present model. This is due to the extreme Ising anisotropy and the one-dimensionality of transverse spin fluctuations  
in the magnetically ordered phase. 

At the temperature where the magnetic phase transition becomes first order we also observe a crossover on the QSL side from a fractionalised paramagnet at low temperatures with exponentially 
suppressed flux excitations to a conventional paramagnet at high temperatures. As expected, the crossover temperature scale, which we identify through a peak in the specific heat,  is set by the energy gap $\Delta$ of flux excitations,
which is equal to $\Delta/K\approx 0.26$ for the isotropic Kitaev model \cite{Kitaev06} and slightly split into $\Delta_z$ and $\Delta_\perp$ by small anisotropy and the Ising exchange. This splitting leads to a broadening of the crossover.  

Interestingly, quantum Monte-Carlo (QMC) simulations of (anisotropic) Kitaev models show a two step thermalization of the QSL state, identified by two 
clearly separated specific heat peaks at temperatures  $T_L$ and $T_H$ \cite{Nasu+14,Nasu+15,Motome+20}. This is in contrast with the single crossover we found within our mean-field treatment. 
For the isotropic Kitaev model the peaks are found at $T_L/K \approx 0.012$  and $T_H/K\approx 0.37$ \cite{Motome+20}. Neither of the crossover temperatures is close to the flux gap 
$\Delta/K\approx 0.26$ \cite{Kitaev06} of the isotropic Kitaev model.  The authors identify the lower temperature peak at $T_L$ with the flux gap and attribute $T_H$ to a feature in the density of states of the 
 itinerant Majorana fermions. However, at $T_H$ the entropy per spin drops from $\ln 2$ to $\frac12 \ln 2$ and the thermal average $\langle \hat W_p\rangle$ of the plaquette operator becomes non-zero, 
suggesting that flux excitations start to freeze out at the temperature $T_H$. The reason why we don't see a crossover at the much lower temperature $T_L$ is likely because at mean-field level the local constraints on the Majorana fermions 
are only treated on average and  the interaction vertex is not properly taken into account. Such correlation effects could give rise to the formation of a bound state at this new energy scale. The inclusion of diagrammatic corrections beyond mean-field 
could potentially provide an analytical confirmation of the QMC result.

The main purpose of our work was to compare different Majorana fermion mean-field theories for Kitaev QSLs. In order to ensure locality of the Hamiltonian after JWT we focused on a very specific, fine tuned spin model. 
A similar Kitaev-Ising model, but with ferromagnetic exchange couplings,  was studied in Ref.~\cite{Nasu+17} for the same reasons, e.g. to ensure locality after a two-dimensional JWT.  
Interestingly, in the regime of strong anisotropy of the Kitaev couplings this model exhibits a spin-nematic phase in between the gapped Kitaev QSL and the ferromagnetic phase. 

It is important to stress that the parton mean-field theory based on the Kitaev mapping to a set of four Majorana fermions, subject to constraints
enforced by Lagrange multipliers, is applicable to a much wider class of extended Kitaev-Heisenberg models \cite{Knolle+18b,Gohlke+17,Burnell+11,schaffer+12,Jiang+11,Reuther+11,Shinjo+15,Joshi18,Smit+20,Consoli+20,Nanda+21}, 
including those relevant to real materials \cite{Singh+10,Liu+11,Singh+12,Choi+12,Ye+12,Singh+12,Takayama+15,Modic+14,Plumb+14,Banerjee+15,Banerjee+17}. 

\section*{Acknowledgments}
F.K. acknowledges fruitful discussions with Zohar Nussinov at the Zaanen Fest, J.L. benefitted from discussions with Sam Carr, Chris Hooley, Aitor Garcia-Ruiz, and Olin Robinson during the Physics by the Lake school. The 
authors thank Joji Nasu for bringing their attention to Ref. \cite{Nasu+17}.

\end{document}